\RequirePackage{fix-cm}
\documentclass[12pt, a4paper, authoryear]{article}
\usepackage[titletoc]{appendix}

\usepackage{xr}
\usepackage{etoolbox}
\usepackage{authblk}
\patchcmd{\emailauthor}{(#2)}{}{}{}

\usepackage[dvipsnames]{xcolor}
\usepackage{graphicx}
\usepackage{amsmath}
\usepackage{amsfonts}
\usepackage{amsthm}
\usepackage{mathtools}
\usepackage{upgreek}
\usepackage{leftidx}
\usepackage{amssymb}
\usepackage{newfile}
\usepackage{endnotes}

\usepackage{booktabs}
\usepackage{multirow}
\usepackage{tabularx}
\usepackage{geometry}
\usepackage{moresize}
\usepackage[hang,flushmargin]{footmisc}
\usepackage{caption}
\usepackage{soul}
\usepackage{paralist}
\usepackage{setspace}
\usepackage{subcaption}
\usepackage{epstopdf}
\usepackage{ragged2e}
\usepackage{floatrow}
\usepackage{hyperref}

\usepackage{pgfplots}
\pgfplotsset{compat=newest}
\usepgfplotslibrary{groupplots}
\usepgfplotslibrary{polar}
\usepgfplotslibrary{smithchart}
\usepgfplotslibrary{statistics}
\usepgfplotslibrary{dateplot}

\usepackage[X2,T1]{fontenc}

\DeclareSymbolFont{cyrillic}{X2}{cmr}{m}{n}
\SetSymbolFont{cyrillic}{bold}{X2}{cmr}{bx}{n}
\DeclareMathSymbol{\CyLje}{\mathord}{cyrillic}{138}
\DeclareMathSymbol{\CyIe}{\mathord}{cyrillic}{170}
\DeclareMathSymbol{\CyBe}{\mathord}{cyrillic}{193}
\DeclareMathSymbol{\CyDe}{\mathord}{cyrillic}{196}
\DeclareMathSymbol{\CyZhe}{\mathord}{cyrillic}{198}
\DeclareMathSymbol{\CyVarZhe}{\mathord}{cyrillic}{199}
\DeclareMathSymbol{\CyI}{\mathord}{cyrillic}{200}
\DeclareMathSymbol{\CyEl}{\mathord}{cyrillic}{203}
\DeclareMathSymbol{\CyTse}{\mathord}{cyrillic}{214}
\DeclareMathSymbol{\CyChe}{\mathord}{cyrillic}{215}
\DeclareMathSymbol{\CySha}{\mathord}{cyrillic}{216}
\DeclareMathSymbol{\CyShcha}{\mathord}{cyrillic}{217}
\DeclareMathSymbol{\CyYer}{\mathord}{cyrillic}{218}
\DeclareMathSymbol{\CyFrontYer}{\mathord}{cyrillic}{220}
\DeclareMathSymbol{\CyE}{\mathord}{cyrillic}{221}
\DeclareMathSymbol{\CyYu}{\mathord}{cyrillic}{222}

\usepackage{lmodern}
\rmfamily
\DeclareFontShape{T1}{lmr}{b}{sc}{<->ssub*cmr/bx/sc}{}
\DeclareFontShape{T1}{lmr}{bx}{sc}{<->ssub*cmr/bx/sc}{}
\usepackage[cal=cm, scr=boondox, frak=euler]{mathalfa} % mathcal

\usepackage [english]{babel}
\usepackage{natbib}
\usepackage[ruled]{algorithm2e}
\usepackage[small]{titlesec}

\setcitestyle{round}

\hypersetup{
	colorlinks = true,
	citecolor = blue,
	linkcolor = blue  
}

\usepackage[nameinlink, noabbrev]{cleveref}

\AtBeginEnvironment{appendices}{\crefalias{section}{appendix}}

\crefrangelabelformat{figure}{#3#1#4--#5#2#6}

\creflabelformat{equation}{#2#1#3}
\crefrangelabelformat{equation}{#3#1#4--#5#2#6}


\titlelabel{\thetitle.\quad}

\newcolumntype{Y}{>{\arraybackslash}X}
\newcolumntype{Z}{>{\centering \arraybackslash}X}

\theoremstyle{plain}

\newtheorem{assumption}{Assumption}
\newtheorem{proposition}{Proposition}

\newtheorem{conjecture}{Conjecture}

\newtheorem{corollary}[proposition]{Corollary}

\theoremstyle{remark}
\newtheorem{definition}{Definition}

\newtheorem{remark}{Remark}

\crefname{lemma}{lemma}{lemmas}
\creflabelformat{lemma}{#2#1#3}
\crefrangelabelformat{lemma}{#3#1#4--#5#2#6}
\crefname{assumption}{assumption}{assumptions}
\creflabelformat{assumption}{#2#1#3}
\crefrangelabelformat{assumption}{#3#1#4--#5#2#6}
\crefrangelabelformat{definition}{#3#1#4--#5#2#6}

\creflabelformat{equation}{#2\textup{#1}#3}
\crefrangelabelformat{algorithm}{#3#1#4--#5#2#6}
\crefrangelabelformat{equation}{#3#1#4--#5#2#6}
\crefrangelabelformat{figure}{#3#1#4--#5#2#6}
\crefrangelabelformat{table}{#3#1#4--#5#2#6}

\newcolumntype{Y}{>{\arraybackslash}X}
\newcolumntype{Z}{>{\centering \arraybackslash}X}

\newcommand{\vect}[1]{\boldsymbol{\mathbf{#1}}}

\newcommand{\expect}{\mathbb{E}}

\newcommand{\defeq}{\vcentcolon=}
\DeclareMathOperator*{\argmax}{arg\,max}
\DeclareMathOperator*{\argmin}{arg\,min}

\DeclareMathOperator*{\cov}{cov}

\DeclareMathOperator{\Tr}{Tr}
\def\indicator{\mathbb{I}}

\newcommand{\widesim}[2][1.5]{
	\mathrel{\overset{#2}{\scalebox{#1}[1]{$\sim$}}}
}

\newcommand{\blpar}{\boldsymbol{(}}
\newcommand{\brpar}{\boldsymbol{)}}

\newcommand\blfootnote[1]{%
	\begingroup
	\renewcommand\thefootnote{}\footnote{#1}%
	\addtocounter{footnote}{-1}%
	\endgroup
}

\title{\fontsize{18}{18} \textsc{\textbf{selecting time-series hyperparameters with the artificial jackknife}}}
\date{}
\author[ ]{Filippo Pellegrino}
\affil[ ]{\it\small Imperial College London}
\affil[ ]{\texttt{f.pellegrino22@imperial.ac.uk}}

\begin{document}

\def\spacingset#1{\renewcommand{\baselinestretch}%
	{#1}\small\normalsize} \spacingset{1}

\spacingset{1.8} % DON'T change the spacing!

% cover page file
\maketitle
\blfootnote{\hspace{1em} I thank Matteo Barigozzi for his valuable suggestions and supervision; Paolo Andreini, Yining Chen, Gianluca Giudice, Thomas Hasenzagl, Cosimo Izzo, Serena Lariccia, Chiara Perricone, Xinghao Qiao, Lucrezia Reichlin, Ragvir Sabharwal, Qiwei Yao, the 2022 IAAE Annual Conference and LSE Workshop on Data Science Theory and Practice participants for their helpful comments on a preliminary draft of this article.}

\begin{abstract}
	This article proposes a generalisation of the delete-$d$ jackknife to solve hyperparameter selection problems for time series. I call it artificial delete-$d$ jackknife to stress that this approach substitutes the classic removal step with a fictitious deletion, wherein observed datapoints are replaced with artificial missing values. This procedure keeps the data order intact and allows plain compatibility with time series. This manuscript justifies the use of this approach asymptotically and shows its finite-sample advantages through simulation studies. Besides, this article describes its real-world advantages by regulating forecasting models for foreign exchange rates.
\end{abstract}

\vspace{0.5cm}

\vfill \null
\setcounter{page}{1}
\clearpage

%%%%%%%%%%%%%%%%%%%%%%%%%%%%%%%%%%%%%%%%%%%%%%%%%%%%%%%%%%%%%%%%%%%%%%%%%
%%%% Main text entry area:
%%%%%%%%%%%%%%%%%%%%%%%%%%%%%%%%%%%%%%%%%%%%%%%%%%%%%%%%%%%%%%%%%%%%%%%%%

% Main

% Notes at the end
%\let\footnote=\endnote
%\patchcmd{\enoteformat}{1.8em}{0pt}{}{}

\section{Introduction}
Using large datasets with standard predictive models is not straightforward. There is often a proliferation of parameters, high estimation uncertainty and the tendency of over-fitting in-sample, but performing poorly out-of-sample. This so-called curse of dimensionality is often handled regularising statistical models with a collection of tuning parameters. Since the latter are often determined before the estimation process takes place, they are denoted as hyperparameters. This paper proposes a systematic approach for selecting them in the case of time-series data.

There is a large number of techniques for prediction problems. Classical methods include ridge \citep{hoerl1970ridge}, LASSO \citep{tibshirani1996regression} and elastic-net \citep{zou2005regularization} regressions. They make the estimation feasible for linear regressions by penalising the magnitude of the coefficients to downweight the variables that do not help in predicting. The strength of the penalties is tuned with a vector of hyperparameters. This manuscript uses the notation $\vect{\gamma}$ to denote relevant vectors of hyperparameters. Regression trees \citep{morgan1963problems, breiman1984classification, quinlan1986induction} are a classical example from the machine learning literature. These techniques can handle non-linearities and complex data generating processes. However, they must be regulated via a range of penalties and stopping rules to perform well out-of-sample. This is again achieved using hyperparameters -- e.g., $\vect{\gamma} = (\text{max depth},\; \text{min number of observations per leaf})$. Large datasets are also commonly handled with Bayesian methods. In this literature, hyperparameters are often necessary to define prior distributions \citep{gelman2014bayesian} and obtain parsimonious models with shrinkage techniques similar or equivalent to ridge and LASSO \citep{giannone2017economic}, and the elastic-net \citep{li2010bayesian}. Hyper-parameters are also crucial for classical problems. For example, hyperparameters such as the number of lags for autoregressive models are fundamental for structuring forecasting exercises -- e.g., $\vect{\gamma} = (\text{lags})$.

Cross-validation \citep{stone1974cross} is among the most well-known approaches for selecting hyperparameters in independent data settings. It is a statistical method to estimate the expected accuracy of a model on unseen data. Its basic formulation is straightforward: data is split into complementary partitions, and the resulting subsamples are used for estimating and validating a predictive method. The performance within the validation samples is used as an estimate of the prediction error on unseen datapoints and the hyperparameters are generally selected to minimise this measure.

Cross-validation is challenging for time series since data is ordered and autocorrelated. Several authors have proposed generalisations to handle these complexities. One of the first contributions came from \cite{snijders1988cross}. The latter used insights from \cite{brown1975techniques} and \cite{ljung1983theory} to propose a cross-validatory method based on realised pseudo out-of-sample errors. Indeed, it suggested to split the observed data into complementary partitions and then use the first as an estimation sample, and the remaining observations to measure the realised pseudo out-of-sample error. The hyperparameters are selected to minimise this error measure.

While this approach is very intuitive and consistent with the structure of the data, it is not necessarily robust, since it uses only a single estimation and validation set. \cite{kunst2008cross} proposed overcoming this downside by applying standard pseudo out-of-sample evaluations to random subsamples. However, the results are relatively difficult to interpret since the algorithm used for generating these partitions is initialised with in-sample regression parameters.

\cite{burman1994cross} introduced a different way to address this problem: the so-called $h$-block cross-validation. This methodology, based on \cite{gyorfi1989nonparametric} and \cite{burman1992data}, uses blocking techniques to generate validation samples independent from the data used for estimation. Indeed, \cite{burman1994cross} proposed creating a set of estimation samples by removing, in turn, each block of dimension $2h+1$ (for a given $h$) from the data. $h$-block cross-validation then uses the median item of this block as a one-dimensional validation sample. Even though this approach has interesting properties, keeping a fixed distance between the partitions is costly, given that a large share of observations is lost in the process. This is especially severe when there are not so many observations, because the number of validation samples available is small.

Most recently, \cite{bergmeir2018note} proposed cross-validating autoregressive models with uncorrelated errors with techniques for i.i.d. data. This approach makes good use of all available observations, but its properties do not hold for models with correlated errors and it disregards the order in the data.

Jackknife \citep{quenouille1956notes, tukey1958bias} and bootstrap \citep{efron1979bootstrap, efron1979computers, efron1981nonparametric} can be used as alternative approaches to estimate the prediction error on unseen datapoints and thus select hyperparameters. These techniques are typically more efficient than cross-validation \citep{efron1979bootstrap, efron1983leisurely} since they measure the accuracy of a model on the average prediction error committed over a large range of data subsamples. Bootstrap builds these partitions sampling with replacement from the data. Instead, jackknife constructs subsamples by removing sets of observations from the observables. In particular, the delete-$d$ jackknife \citep{wu1986jackknife, shao1989general} generates a sequence of partitions by removing, in turn, all the combinations of $d>0$ observations from the data.

Jackknife and bootstrap require modifications to be compatible with time series, since the subsampling schemes do not take the data order into account. \cite{kunsch1989jackknife} extended these methodologies to stationary series. Indeed, building on \cite{carlstein1986use}, \cite{kunsch1989jackknife} proposed developing block-wise subsampling schemes. Let $c$ be an integer lower or equal to the total number of observed time periods. The block jackknife generates the partitions by removing or down-weighting, in turn, all the $c$-dimensional blocks of consecutive observations from the data. Instead, the block bootstrap draws with replacement a fixed number of $c$-dimensional blocks of observations from the data. \cite{politis1992circular, politis1994stationary} developed this technique further proposing the so-called stationary bootstrap. This approach wraps the data ``in a circle'', so that the first observation follows the last, and generates the bootstrap samples drawing and merging blocks of random length. Differently than the block bootstrap and its variations, the block jackknife does not impact the data order when constructing subsamples.

	\begin{figure}[t!]
		\centering
		\begin{subfigure}[b]{0.475\textwidth}
			\centering
			\includegraphics[height=3.5cm, width=6.5cm]{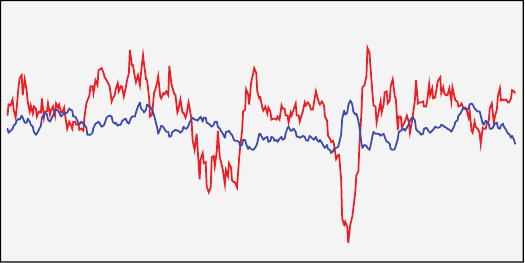}
			\caption{Observed stationary time series.} 
		\end{subfigure}
		\hfill
		\begin{subfigure}[b]{0.475\textwidth}  
			\centering 
			\includegraphics[height=3.5cm]{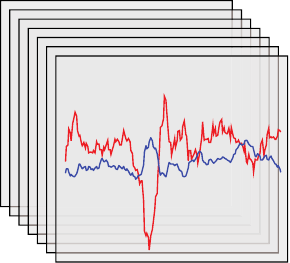}
			\caption{Bootstrap subsamples.}
		\end{subfigure}
		\vskip\baselineskip\vspace{15pt}
		\begin{subfigure}[b]{0.475\textwidth}   
			\centering 
			\includegraphics[height=3.5cm, width=6.5cm]{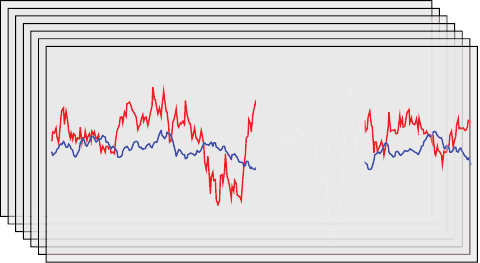}
			\caption{Block jackknife subsamples.}  
		\end{subfigure}
		\hfill
		\begin{subfigure}[b]{0.475\textwidth}   
			\centering 
			\includegraphics[height=3.5cm, width=6.5cm]{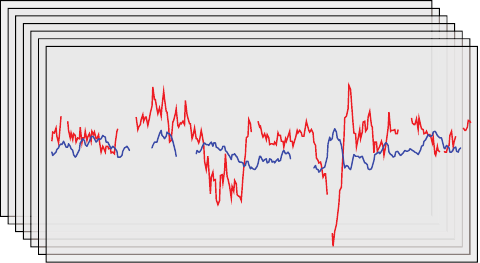}
			\caption{Artificial delete-$d$ jackknife subsamples.}
		\end{subfigure}
		\vspace{5pt}
		\caption{Subsampling schemes for dependent data.}
		\floatfoot{\justifying (b) Blocks of random (stationary bootstrap) or fixed (block bootstrap) length are drawn with replacement from the data. (c) Subsamples are constructed down-weighting, in turn, all the $c$-dimensional blocks of consecutive observations from the data. As in \cref{ch1:sec:methodology:existing_estimators}, the down-weighting scheme is operated by turning blocks of consecutive observations into missing values. (d) Subsamples are constructed imposing (artificial) patterns of missing data to the original sample. This is a generalisation of the delete-$d$ jackknife.}
	\end{figure}

This paper introduces a version of the standard delete-$d$ jackknife compatible with time series. In this version of the jackknife, the data removal step is replaced with a fictitious deletion that consists in imposing (artificial) patterns of missing observations on the data. I call this new approach artificial delete-$d$ jackknife (or artificial jackknife) to emphasise that $d$ observations are artificially removed from the original data to generate each subsample. This article proposes using this new methodology to compute a robust measure of the forecast error (or, artificial jackknife error) as a means for selecting hyperparameters. The advantages of this approach depend on the finite-sample properties of the artificial jackknife. In fact, all errors based on pseudo out-of-sample evaluations converge in probability to the true error with the same rate (as shown in \cref{ch1:appendix:asymptotics}). However, the artificial jackknife error has a smaller finite-sample variance than the pseudo out-of-sample error and the block jackknife (for most configurations of $c$ and $d$). This is crucial for stability and to select hyperparameters when the number of observations (i.e., time periods) is limited.

The artificial delete-$d$ jackknife is compatible with forecasting models able to handle missing observations. Within the scope of this paper, this is not a strong restriction. Most predictive problems with missing observations in the measurements can be written in state-space form and estimated via a large number of methods, as surveyed in \citet[ch. 6]{shumway2011time} and \citet[ch. 12]{sarkka2013bayesian}. 

As an illustration, this article employs the artificial jackknife for tuning penalised vector autoregressive moving average (VARMA) models. In this case, the vector of hyperparameters $\vect{\gamma}$ includes traditional elastic-net tuning parameters \citep{zou2005regularization}. These models are estimated on a dataset of weekly exchange rate returns. In order to provide full compatibility with the artificial jackknife, this article proposes to estimate the VARMAs with an Expectation-Conditional Maximisation (ECM) algorithm \citep{meng1993maximum} able to handle incomplete time series. This estimation method is a secondary contribution of the paper given that, to my best knowledge, the literature has not proposed a way for handling missing observations in the measurements with similar settings.\footnote{The replication code for this empirical application is available on \href{https://github.com/fipelle/replication-pellegrino-2022-hyperparameters}{GitHub}.}

\section{Methodology}\label{ch1:sec:methodology}

One of the main objectives of time series is to predict the future. This article aims to select optimal vectors of hyperparameters consistently with this maxim and thus in a way that minimises the expected forecast error.

\subsection{Foundations}\label{ch1:sec:methodology:foundations}

This subsection sets out the foundations for the hyperparameter selection process and delimits the scope of the article to a broad family of forecasting methods. Note that this subsection and \cref{ch1:sec:methodology} in its entirety do not limit the manuscript by looking at a specific forecasting model. Hence, the theoretical results are widely applicable. 

\begin{assumption}[Data] \label{ch1:assumption:data}
	Let $n, T \in \mathbb{N}$ and $n_{Z} \in \mathbb{N}_{0}$. Assume that $Y_{i,t}$ and $Z_{j,t}$ are finite realisations of some real-valued stochastic processes observed at time periods in the sets $\mathscr{T}_{i}, \mathscr{T}_{j} \subseteq \{t : t \in \mathbb{Z},\, 1 \leq t \leq T\}$ for $i=1, \ldots, n$ and $j=n+1, \ldots, n+n_Z$. 
\end{assumption}

\begin{assumption}[Lags] \label{ch1:assumption:lags}
	Define $q, r \in \mathbb{N}_{0}$ to be such that $p \defeq \max(q,r)$ and $0 < p \ll T-1$. 
\end{assumption}

\begin{assumption}[Predictors] \label{ch1:assumption:predictors}
	 Let $\vect{X}_t \defeq (\vect{Y}_{t}' \; \ldots \vect{Y}_{t-q+1}'\; \vect{Z}_{t}' \; \ldots \vect{Z}_{t-r+1}')'$ be $m \times 1$ and defined at any point in time $t \in \mathbb{Z}$. 
\end{assumption}

\begin{assumption}[Model structure] \label{ch1:assumption:structure} 
	Finally, assume that
	\begin{align}
	\vect{Y}_{t+1} = \vect{f} (\vect{X}_{t}, \vect{\Psi}) + \vect{V}_{t+1}, \label{ch1:eqautoregressive}
	\end{align}
	
	\noindent where $\expect \left[ \Vert \vect{f} (\vect{X}_{t}, \vect{\Psi}) \Vert_{2}^2 \right] < \infty$, $\vect{\Psi}$ is a matrix of finite coefficients, $\vect{V}_{t+1} \widesim{i.i.d.} \left(\vect{0}_{n \times 1}, \vect{\Sigma} \right)$ with $\vect{\Sigma}$ being a positive definite matrix\footnote{This part of \cref{ch1:assumption:structure} could be relaxed following an approach similar to the one employed in \cite{barigozzi2020quasi}. However, this is outside the scope of the paper.} and $\expect(\vect{V}_{t+1} | \vect{X}_t) = 0$, for any integer $t$. Under \crefrange{ch1:assumption:data}{ch1:assumption:predictors}, $\vect{X}_t$ is allowed to include $\vect{Y}_{t}, \ldots, \vect{Y}_{t-q+1}$ and $\vect{V}_{t}, \ldots, \vect{V}_{t-r+1}$ (for some $0 \leq q \leq p$ and $0 \leq r \leq p$), and more explanatory variables referring up to time $t-p+1$.
\end{assumption}

\begin{remark}
	The dependence on the sample size is highlighted in the notation only when strictly necessary, in order to ease the reading experience. Also, this article uses the same symbols to indicate the realisations at some integer point in time $t$ and their general value in the underlying process. This is again for simplifying the notation and it should be clear from the context whether the manuscript is referring to the first or second category.
\end{remark}

\begin{remark}
	The lags $q$ and $r$ are also assumed to be known. This might be considered a strong assumption, as they can be viewed as hyperparameters. However, if $q$ and $r$ are set to be large, the selection of lags can equivalently be seen as a shrinkage problem concerning the coefficients. A related approach is employed for the empirics.
\end{remark}

\begin{remark}
	Assumption 4 encompasses stationary and a subset of non-stationary cases. Specifically, this subset includes non-stationary cases where the functional form of $\vect{f}(\cdot)$ remains unchanged over time.This is the case, for instance, of vector autoregression processes with a unit root and fixed parameters. Non-stationary cases with time-varying $\vect{f}(\cdot)$ are beyond the scope of this paper.
\end{remark}

Knowing the data generating process in \cref{ch1:assumption:structure}, one could use it for obtaining the most accurate prediction (true forecast) for $\vect{Y}_{t+1}$ given $\vect{X}_t$ at any point in time $t$. 

\begin{definition}[True error]\label{ch1:def:trueerror}
	Under a weighted square loss, the expected error associated with the true forecast is
	\begin{align*}
		err \defeq \sum_{i=1}^n w_i \, \expect \left[\left|Y_{i, t+1} - f_i \left(\vect{X}_{t}, \vect{\Psi}\right)\right|^2\right] = \sum_{i=1}^n w_i \, \expect (V_{i, t+1}^2) = \sum_{i=1}^n w_i \, \vect{\Sigma}_{i,i},
	\end{align*}
	
	\noindent with $w_i \geq 0$ for $1 \leq i \leq n$. This article refers to $err$ as the true error.
\end{definition}

\noindent In most practical applications, the data generating process is unknown and forecasters' objective can be then reduced to approximating the true forecast.

\begin{definition}[Information set]
	At any time period $p \leq s \leq T$, forecasters have an information set $\mathscr{I}(s)$ containing the data observed up to that point.
\end{definition}

\begin{definition}[Forecast function]
	Let $\vect{g}$ be a function that maps a vector of covariates \(\vect{X}_t\) and a set of parameters \(\vect{\hat{\theta}}_s(\vect{\gamma})\) into forecasts with $\expect \left[ \Vert \vect{g} \blpar \vect{X}_{t}, \vect{\hat{\theta}}_{s} (\vect{\gamma}) \brpar \Vert_{2}^2 \right] < \infty$. The coefficients \(\vect{\hat{\theta}}_{s}(\vect{\gamma})\) are estimated on the basis of the data in $\mathscr{I}(s)$, and given a vector of hyperparameters $\vect{\gamma}$.
\end{definition}
	
\begin{assumption}[Forecasts conditional on information set] 
	\label{ch1:assumption:infset}
	For each time \(t\) with \(p \le s \le T\), the forecaster's expectation for $\vect{Y}_{t+1}$ conditional on $\mathscr{I}(s)$ is
	\begin{align*}
		\vect{\hat{Y}}_{t+1|s}(\vect{\gamma}) \defeq \expect \left[\vect{Y}_{t+1} | \vect{X}_t, \vect{\hat{\theta}}_{s} (\vect{\gamma}) \right] = \vect{g} \blpar \vect{X}_{t}, \vect{\hat{\theta}}_{s} (\vect{\gamma}) \brpar,
	\end{align*}
\end{assumption}

\begin{remark}
	The forecast function is further specified in \cref{ch1:sec:methodology:asymptotics} with \cref{ch1:assumption:mse}. Note that forecasters' may consider different predictors than those in the true forecast function when constructing their approximation. The article does not explicitly consider this case to simplify notation, but allowing for it would not change the results.
\end{remark}

\noindent The predictions generated as in \cref{ch1:assumption:infset} are clearly less accurate than the corresponding true forecasts. However, as shown in the following subsections, empirical error estimators converge in probability to the true error for a wide class of forecast functions. Therefore, the use of these approximations can be justified asymptotically. 

\subsection{Existing error estimators}\label{ch1:sec:methodology:existing_estimators}

This subsection expands on these empirical error estimators. It starts by describing the most well-known and broadens the discussion with the block jackknife.

Before getting into details, I need to define a loss function for measuring the forecast error at each point in time.

\begin{definition}[Loss]\label{ch1:def:forecasterr}
	
	Consistently with \cref{ch1:def:trueerror}, this paper uses
	\begin{align*}
		L \blpar \vect{Y}_{t+1}, \vect{\hat{Y}}_{t+1|s}(\vect{\gamma}) \brpar \defeq \sum_{i \in \mathscr{D}(t+1)} w_i \, \big[Y_{i,t+1} - \hat{Y}_{i,t+1|s}(\vect{\gamma}) \big]^2,
	\end{align*}
	
	\noindent where $\mathscr{D}(t+1) \defeq \{i : 1\leq i \leq n \text{ and } Y_{i,t+1} \neq \text{NA}\}$, NA denotes a generic missing value, for any $p \leq t \leq T-1$ and $p \leq s \leq T$.
\end{definition}

The next thing to consider is the conceptual relation between the forecast and its conditioning set. As a general point, the difficulty in obtaining an accurate $\vect{\hat{Y}}_{t+1|s}(\vect{\gamma})$ changes depending on whether $\mathscr{I}(s)$ includes information about the future. This is what leads to the distinction between the two most common categories of forecast error estimators: in-sample and pseudo out-of-sample.

\begin{definition}[In-sample error] \label{ch1:def:iiserror}
	The in-sample error
	\begin{align*}
	\overline{err}(\vect{\gamma}) \defeq \frac{1}{T-p}\sum_{t=p+1}^{T} L \blpar \vect{Y}_t, \vect{\hat{Y}}_{t|T}(\vect{\gamma}) \brpar
	\end{align*}

	\noindent is a measure of the average loss between the data and predictions generated conditioning on the full information set. 
\end{definition}

\noindent Estimating the coefficients once and on the full information set is beneficial for very short time-series problems, as there may not be enough observations to compute more sophisticated estimators. However, this approach tends to overstate the forecast accuracy since the information set is (at least partially) aware of the future. Indeed, in a realistic environment, forecasters would only have information about the past when computing their predictions.

\begin{definition}[Pseudo out-of-sample error]\label{ch1:def:ooserror}
	The pseudo out-of-sample error
	\begin{align*}
	\widehat{err}(\vect{\gamma}) \defeq \frac{1}{T-t_0} \sum_{t=t_0}^{T-1} L \blpar \vect{Y}_{t+1}, \vect{\hat{Y}}_{t+1|t}(\vect{\gamma}) \brpar,
	\end{align*}
	
	\noindent overcomes this limitation by using forecasts generated on the basis of an expanding and backward-looking information set, starting from $p \leq t_0 \leq T-1$.
\end{definition}

\begin{remark}
	The pseudo out-of-sample error can be extended to forecast horizons larger than one, but this is not further explored in the manuscript. When long run predictions are calculated iteratively from the one step ahead forecast, it is not necessary to generalise \cref{ch1:def:ooserror} to handle longer horizons. The latter would need to be modified only in the case of direct forecast. It is important to stress that when the model is correctly specified, producing iterative forecasts is more efficient than computing horizon-specific ones. However, the latter are more robust to misspecification \citep{marcellino2006comparison}.  For simplicity, this paper focusses only on the one-step ahead forecast, wherein iterative and direct forecasts are identical. Implicitly, this approach is also consistent with iterative forecast methods targetting longer horizons.
\end{remark}

Unfortunately, the pseudo out-of-sample error can be either over or under confident depending on the time periods used for estimating and validating the model. This can be overcome using estimators based on the average of pseudo out-of-sample errors computed on a series of data subsamples. The article generates these partitions using time-series generalisations of the jackknife \citep{quenouille1956notes, tukey1958bias}. The generic jackknife error in \cref{ch1:def:jackerror} is an estimator that can accommodate for different jackknife partitioning algorithms.

\begin{definition}[Generic jackknife error]\label{ch1:def:jackerror}
	Let $\mathscr{J}$ be an indexed family of sets such that each element contains ordered pairs $(i,t)$ with $1 \leq i \leq n$ and $0 \leq t \leq T$. The generic jackknife pseudo out-of-sample error
	\begin{align*}
	\widetilde{err}(\mathscr{J}, \vect{\gamma}) \defeq \frac{1}{|\mathscr{J}| \cdot (T-t_0)} \sum_{j=1}^{|\mathscr{J}|} \sum_{t=t_0}^{T-1}  L \blpar \vect{Y}^{-j}_{t+1}, \vect{\hat{Y}}^{-j}_{t+1|t}(\vect{\gamma}) \brpar, %\label{ch1:eqgenericjkerr}
	\end{align*}
	
	\noindent where $\vect{Y}^{-j}$ is the $n \times T$ matrix such that
	\begin{alignat*}{2}
	Y^{-j}_{i,t} &\defeq \begin{cases}
	Y_{i,t}, &\text{ if } (i,t) \notin \mathscr{J}_j, \\
	\text{NA}, &\text{ if } (i,t) \in \mathscr{J}_j,
	\end{cases}
	\end{alignat*}
	
	\noindent $\vect{\hat{Y}}^{-j}_{t+1|t}(\vect{\gamma})$ is analogous to $\vect{\hat{Y}}_{t+1|t}(\vect{\gamma})$, but the autoregressive data component of $\vect{X}_{t}$ is now based on $\vect{\hat{Y}}^{-j}$. As for \cref{ch1:def:ooserror}, $p \leq t_0 \leq T-1$.
\end{definition}

\begin{remark}
	Allowing the ordered pairs $(i,t)$ to have $t=0$, permits to write the pseudo out-of-sample error as a banal case of jackknife error in which $\mathscr{J}$ contains only one element external to the sample. For instance via $\mathscr{J}=\{(1,0)\}$. Indeed, the actual data has observations referring to the points in time between $1$ and $T$ (included). Therefore, any $t=0$ is to be considered external.
\end{remark}

The most well-known approach to generate jackknife subsamples for dependent data is the block jackknife \citep{kunsch1989jackknife}. This technique partitions the data into block jackknife samples by removing or down-weighting, in turn, all the unique non-interrupted blocks of $1 \leq c \leq T$ observations.

\begin{definition}[Block jackknife error]\label{ch1:def:bjkerror}

	This paper denotes the block jackknife error as
	\begin{align}
	\widetilde{err}^{BJK}(c, \vect{\gamma}) \equiv \widetilde{err} \blpar \mathscr{B}(c), \vect{\gamma} \brpar, \label{ch1:eqbjkoos}
	\end{align}
	
	\noindent where $\mathscr{B}(c)$ is the family of sets
	\begin{align*}
	&\mathscr{B}(c) \defeq \{ \mathscr{B}(1, c), \ldots,  \mathscr{B}(T-c+1, c) \}
	\end{align*}
	
	\noindent and
	\begin{align*}
	&\mathscr{B}(j, c) \defeq \{(i,t): 1 \leq i \leq n \text{ and } j \leq t \leq j+c-1 \}.
	\end{align*}
\end{definition}

\begin{remark}
	\noindent In other words, this article constructs the individual blocks by replacing, in turn, all the unique non-interrupted blocks of $c$ observations with missing values. This is compatible with \cite{kunsch1989jackknife} since imposing blocks of NAs can be interpreted as fully down-weighting groups of observations. Furthermore, it simplifies the use of the block jackknife to estimate hyperparameters in forecasting settings. By processing the data via filtering and smoothing methods compatible with missing observations, it is easier to estimate forecasting models without pre-processing the measurements to remove breaks introduced in the subsampling process.
\end{remark}

\subsection{A novel error estimator}\label{ch1:sec:methodology:novel_estimators}

Building on the discussion of empirical error estimators, this subsection takes a closer look at jackknife-based methods and introduces a novel version: the artificial delete-$d$ jackknife error estimator.

The main issue with the block jackknife error is that the number of partitions that can be generated from the data is generally small. Thus, the overall improvement over the standard pseudo out-of-sample error is somewhat limited. Also, for those partitions wherein a huge chunk of observations are removed after $t_0$, dividing for a factor of $T-t_0$ may produce inaccurate estimates of the expected error. This is especially true in small-sample problems where $c$ is large relative to $T-t_0$. A simple way for reducing this issue in a finite-sample problem consists in adjusting the $\widetilde{err}^{BJK}(c, \vect{\gamma})$ multiplying it by
\begin{align*}
\frac{T-t_0}{|\mathscr{B}(c)|} \sum_{j=1}^{|\mathscr{B}(c)|} \frac{1}{|\{(i,t) \in \mathscr{B}(j,c) : t > t_0\}|}.
\end{align*}

\noindent However, this is difficult to justify asymptotically. 

This paper proposes to surpass these problems using an error estimator based on a generalisation of delete-$d$ jackknife \citep{wu1986jackknife, shao1989general} compatible with time-series problems: the artificial delete-$d$ jackknife. The classical delete-$d$ jackknife for i.i.d. data \citep{wu1986jackknife, shao1989general} generates subsamples by removing, in turn, all the combinations of $d > 0$ observations from the data. This is clearly incompatible with dependent data, since the autocorrelation structure would break during the subsampling process. The artificial jackknife overcomes this complexity by generating the partitions replacing, in turn, all the combinations of $d$ observations with (artificial) missing values. This allows to handle dependent data, as the resulting partitions keep the original ordering and the autocorrelation structure is not altered. Moreover, this approach permits to generate a much larger number of subsamples than block jackknife.\footnote{It is interesting to notice that the block jackknife in \cref{ch1:eqbjkoos} is a special case of the artificial delete-$d$ jackknife, in which blocks of consecutive datapoints are replaced with missing values.}

\begin{definition}[Artificial delete-$d$ jackknife error]\label{ch1:def:ajkerror}
	Let
	\begin{align*}
	\mathscr{P} \defeq \{i \in \mathbb{Z}: 1 \leq i \leq n \} \times \{t \in \mathbb{Z}: 1 \leq t \leq T \}
	\end{align*}
	
	\noindent be the set of all data pairs. Hence, define $\mathscr{A}(d)$ as a family of sets with cardinality
	\begin{align*}
	|\mathscr{A}(d)| = \frac{(nT)!}{d!\, (nT-d)!}
	\end{align*}
	
	\noindent such that each element is a $d$-dimensional combination of $\mathscr{P}$. Next, let
	\begin{align}
	\widetilde{err}^{AJK}(d, \vect{\gamma}) \equiv \widetilde{err} \blpar \mathscr{A}(d), \vect{\gamma} \brpar \label{ch1:eqajkoos}
	\end{align}

	\noindent to simplify notation. This is the artificial delete-$d$ jackknife error.
\end{definition}

\noindent The higher reliability of this error estimator is given by the large number of partitions that the artificial delete-$d$ jackknife is able to generate and their heterogeneity. This can be formalised in terms of efficiency as follows.

\begin{assumption}[Finite-sample variance] \label{ch1:assumption:finite_sample_var}
	Assume that the constituent pseudo out-of-sample errors in \cref{ch1:def:jackerror} follow a common finite-sample distribution with variance $\sigma^2(T-t_0, \vect{\gamma})$.
\end{assumption}

\begin{proposition} \label{ch1:proposition:finite_sample}
	Under \cref{ch1:assumption:finite_sample_var}, it follows that, in finite-sample problems, 
	\begin{align*}
		&\text{var}\left[\widetilde{err}^{AJK}(d, \vect{\gamma})\right] \leq \text{var} \blpar \widehat{err}(\vect{\gamma}) \brpar, \\
	&\text{var}\left[\widetilde{err}^{BJK}(c, \vect{\gamma})\right] \leq \text{var} \blpar \widehat{err}(\vect{\gamma}) \brpar.
	\end{align*}
\end{proposition}

\begin{proof}
	The proof is reported in \cref{ch1:appendix:finite_sample:proof}.
\end{proof}

\begin{remark}
	\Cref{ch1:assumption:finite_sample_var} is introduced to simplify the notation used in the proof of \cref{ch1:proposition:finite_sample}, in a manner similar in spirit to \citet[Equation 15.1]{hastie2009elements}. While this assumption is not essential to the overall structure of the proof and can be relaxed, doing so results in significantly more complex notation. An alternative approach, which yields equivalent results, would involve assuming that $\text{var} \left[\widehat{err}^{-i}(\vect{\gamma}) \right]$ is bounded by a finite maximum over $i$.
\end{remark}

\noindent \Cref{ch1:appendix:finite_sample:simulations} compares the variance of the block and artificial jackknife errors through a simulation exercise. This exercise shows that the artificial jackknife outperforms the block jackknife especially in small-sample problems.

When $nT$ is large and $\sqrt{nT} < d < nT$, the cardinality $|\mathscr{A}(d)|$ can be large and it might not be computationally feasible to calculate \cref{ch1:eqajkoos} evaluating all combinations. Following common practice \citep[p. 149]{efron1994introduction}, this computational issue is handled with an approximation. Define $\widetilde{\mathscr{A}}(d) \subset \mathscr{A}(d)$ as a family of sets constructed by drawing at random, without replacement, for a sufficiently large number of times from $\mathscr{A}(d)$. Hence, use this newly defined subset to compute $\widetilde{err} \blpar \widetilde{\mathscr{A}}(d), \vect{\gamma} \brpar$, an approximation of the artificial delete-$d$ jackknife error. Clearly, the accuracy of the approximation
\begin{align}
	\widetilde{err}^{AJK}(d, \vect{\gamma}) \approx \widetilde{err} \blpar \widetilde{\mathscr{A}}(d), \vect{\gamma} \brpar \label{ch1:eqapproxajkoos}
\end{align}
	
\noindent depends on how close $|\widetilde{\mathscr{A}}(d)|$ is to $|\mathscr{A}(d)|$. 

In most empirical problems, the artificial jackknife will likely be truncated. Thus, this article proposes a simple heuristics for selecting its number of artificial missing observations. As detailed in \cref{ch1:appendix:finite_sample} and with the simplified notation in \cref{ch1:assumption:finite_sample_var}, the artificial jackknife error variance depends on two factors: $\sigma^2(T-t_0, \vect{\gamma})$ and the heterogeneity across jackknife subsamples. The latter is controlled by $d$ and, ceteribus paribus, $\text{var} \blpar \widetilde{err}^{AJK}(d, \vect{\gamma}) \brpar$ is at its minimum when the subsamples are the most diverse. The exact functional form of this variance is unknown and, in the case of the truncated artificial jackknife, one would need to choose a value for $d$ that guarantees a large pool of combinations. Besides, it would be ideal to exclude from $\widetilde{\mathscr{A}}(d)$ the combinations that are the most similar to the block jackknife. In other words, those where all series are missing for one or more periods.

\begin{conjecture}[Rule of thumb for selecting d]
	As a result, this paper proposes selecting $d$ for the truncated artificial jackknife error to be
	\begin{align}
		\hat{d} = \argmax_{\underline{d}} \binom{nT}{\underline{d}} - \indicator_{\underline{d} \, \geq \, n} \, \binom{nT - n}{\underline{d} -n} \, T - \sum_{i=2}^{\lfloor \underline{d}/n \rfloor} (-1)^{i-1} \, \binom{T}{i} \, \binom{nT - in}{\underline{d}-in}, \label{ch1:eqdhat}
	\end{align}

	\noindent where
	\begin{align*}
		\indicator_{d \geq n} \, \binom{nT - n}{d -n} \, T + \sum_{i=2}^{\lfloor d/n \rfloor} (-1)^{i-1} \, \binom{T}{i} \, \binom{nT - in}{d-in},
	\end{align*}

	\noindent is the amount of subsamples with points in time where all series are artificially missing.\footnote{To further reduce the effect of those combinations where all series are missing in one or more points in time, the simulation algorithm employed in this article is structured to exclude them from $\widetilde{\mathscr{A}}(d)$.}
\end{conjecture}

\begin{remark}
	The maximisation is trivial since the objective function is particularly fast to compute for each admissible $\underline{d}$, that is every integer $\underline{d} \in [1, nT]$.
\end{remark}

\subsection{Asymptotic properties and hyperparameter selection }\label{ch1:sec:methodology:asymptotics}

This subsection provides the asymptotic justification needed for using the approximation in \cref{ch1:assumption:infset} to forecast the target data. It starts by describing the underlying assumptions and continues by proving that pseudo out-of-sample evaluations are consistent, even in the presence of missing observations. The proofs are reported in \cref{ch1:appendix:asymptotics}. It then concludes showing how to employ these estimators to select vectors of hyperparameters $\vect{\gamma}$. 

\begin{assumption}[Bounds] For any finite $n>0$, 
	\begin{align*}
		&\sum_{i=1}^{n} w_i \leq M_1, \\
		&\sum_{i=1}^{n}  \sum_{j=1}^{n} | \cov (V_{i, t}, V_{j, t}) | \leq M_2, \\
		&\sum_{i=1}^{n} \sum_{j=1}^{n} | \cov ( V_{i, t}^2, V_{j, t}^2) | \leq M_3,
	\end{align*}\label{ch1:assumption:summability}
	
	\noindent where $M_1, M_2, M_3 \in (0, \infty)$ are non-negative finite constants.
\end{assumption}

\begin{remark}
	Recall that the elements of $\vect{w}$ are non-negative. Thus, for any finite $n>0$, 
	\begin{align*}
		\sum_{i=1}^{n} w_i = \sum_{i=1}^{n} | w_i |
	\end{align*}
	\noindent by definition. The case for $n \rightarrow \infty$ is beyond the scope of this article.
\end{remark}

\begin{assumption}[Mean squared error of the forecast] For any $t>0$,
	\begin{align*}
		&\expect \left( \Vert \vect{f} (\vect{X}_{t}, \vect{\Psi}) - \vect{g} \blpar \vect{X}_{t}, \vect{\hat{\theta}}_{t}(\vect{\gamma}) \brpar \Vert_{2}^2 \right) \leq M_4 / t,
	\end{align*}
	
	\noindent where $M_4 \in (0, \infty)$ is a positive finite constant. 

\label{ch1:assumption:mse}
\end{assumption}

\begin{remark}
	Note that
	\begin{align*}
		\sup_{t} \Vert \vect{f} (\vect{X}_{t}, \vect{\Psi}) - \vect{g} \blpar \vect{X}_{t}, \vect{\hat{\theta}}_{t}(\vect{\gamma}) \brpar \Vert_{2}^2 \leq \sup_{t} \Vert |\vect{f} (\vect{X}_{t}, \vect{\Psi})| + |\vect{g} \blpar \vect{X}_{t}, \vect{\hat{\theta}}_{t}(\vect{\gamma}) \brpar| \Vert_{2}^2.
	\end{align*}
	
	\noindent Since under \crefrange{ch1:assumption:structure}{ch1:assumption:infset} both the true forecast and its approximation are always finite, the assumption holds within the context of this paper. However, this bound can be loose as it is a function on the problem at hand and it depends on the true forecast and all modelling choices. 
\end{remark}

\begin{remark}
	The current form of this assumption is less accommodating of more general, potentially non-linear or semi-parametric models. Some may only achieve a slower rate. In that case, one can generalize the assumption to require
	\[
	\mathbb{E}\!\Big[\|\mathbf{f}(\mathbf{X}_{t},\mathbf{\Psi}) \;-\; \mathbf{g}\bigl(\mathbf{X}_{t},\widehat{\boldsymbol{\theta}}_{t}(\gamma)\bigr)\|_{2}^2\Big]
	\;\;\le\;\; \frac{M_{4}}{t^{\alpha}}, 
	\]
	for some \(\alpha \in (0,1]\). Whether \(\alpha = 1\) or \(\alpha < 1\) will depend on the structure of the model and the estimation procedure.
\end{remark}

\begin{assumption}[Limiting size of the presample] Assume that
	\begin{align*}
		&\lim_{T \rightarrow \infty} t_0 / T = 0.
	\end{align*}
	\noindent as $t_0$ goes to infinity. 
	
\label{ch1:assumption:presample}
\end{assumption}

\begin{assumption}[Limiting number of missing observations] Denote with $0 \leq t_{NA} < T-t_0$ the number of periods between $t_0+1$ and $T$ (included) where the data contains missing observations only, and assume that
	\begin{align*}
		\lim_{T \rightarrow \infty} t_{NA} / T = 0, 
	\end{align*}
	\noindent as $t_{NA}$ goes to infinity. 
	
\label{ch1:assumption:missings}
\end{assumption}

\begin{remark}
	Note that \cref{ch1:assumption:presample} serves a crucial purpose: making sure that as $T$ approaches infinity, the pseudo out-of-sample period increases. Similarly, \cref{ch1:assumption:missings} limits the number of periods with missing observations, as $T$ approaches infinity. This implies that as $T$ increases the information set expands, because the number of observed datapoints increases. Without \cref{ch1:assumption:missings}, the total number of missing values could become predominant, relative to the amount of observed datapoints.
\end{remark}

\begin{proposition}\label{ch1:proposition:nomissing}
	Denote with $\widehat{err}_T(\vect{\gamma})$ the pseudo out-of-sample error for a dataset with $T$ periods. Under \crefrange{ch1:assumption:data}{ch1:assumption:infset} and \crefrange{ch1:assumption:summability}{ch1:assumption:presample}, and with complete data it holds that
	\begin{align*}
		\lim_{T \rightarrow \infty} \sqrt{T} \, \expect \bigg[\big|\widehat{err}_T(\vect{\gamma}) - err\big|\bigg] \leq 4M_1\sqrt{M_2}\sqrt{M_4}.
	\end{align*}
\end{proposition}

This proposition shows that with complete data, pseudo out-of-sample errors are consistent estimators of the true error. This is a first stepping stone to prove convergence in probability for the generic jackknife errors. \Cref{ch1:proposition:withmissings} bridges further the gap by extending these results to estimators based on potentially incomplete data.

\begin{proposition}\label{ch1:proposition:withmissings}
	\noindent Under \crefrange{ch1:assumption:data}{ch1:assumption:infset} and \crefrange{ch1:assumption:summability}{ch1:assumption:missings}, and with potentially incomplete data it holds that
	\begin{align*}
		\lim_{T \rightarrow \infty} \sqrt{T} \, \expect \bigg[\big|\widehat{err}_T(\vect{\gamma}) - err\big|\bigg] \leq 4M_1\sqrt{M_2}\sqrt{M_4}.
	\end{align*}	
\end{proposition}

\begin{remark}
	Under \cref{ch1:assumption:missings} the rate of convergence in \cref{ch1:proposition:nomissing} is preserved with potentially incomplete data.
\end{remark}

The following corollary of \cref{ch1:proposition:withmissings} extends its conclusions to the generic jackknife pseudo-out-of-sample error estimators described in \cref{ch1:sec:methodology:existing_estimators}. Clearly, this includes the artificial delete-$d$ jackknife error.

% Corollary
\begin{corollary}\label{ch1:corollary:jackknife}
	Let $\widetilde{err}_T(\mathscr{J}, \vect{\gamma})$ be a generic jackknife pseudo out-of-sample error based on a dataset with $T$ time periods. Under the assumptions of \cref{ch1:proposition:withmissings}, it holds that
	\begin{align*}
		\lim_{T \rightarrow \infty} \sqrt{T} \, \expect \bigg[\big|\widetilde{err}_T(\mathscr{J}, \vect{\gamma}) - err\big|\bigg] \leq 4M_1\sqrt{M_2}\sqrt{M_4}.
	\end{align*}
\end{corollary}

Having justified asymptotically the use of these estimators, this subsection proceeds by showing how to select hyperparameters. It does so by exploring a grid of candidate hyperparameters to find the minimiser for a pseudo out-of-sample error estimator of choice between those reported above. This is in line with classical empirical risk minimisation \cite[see][ch. 3 for a complete survey]{elliot2016economic}. This article builds on \cite{bergstra2012random} and does so via random search. 

\begin{definition}[Random search] \label{ch1:def:randomsearch}
	Let $\mathscr{H}$ be a compact set of ordered tuples that defines the region of existence of the vector of hyperparameters of interest. Define $\mathscr{H}^{RS} \subseteq \mathscr{H}$ as a set of candidate vectors of hyperparameters constructed via means of independent and uniform draws without replacement from $\mathscr{H}$. A random search considers every candidate in $\mathscr{H}^{RS}$ and computes
	\begin{align*}
		\begin{cases}
			\vect{\hat{\gamma}}(\mathscr{H}^{RS}) \defeq \argmin_{\underline{\vect{\gamma}} \in \mathscr{H}^{RS}} \widehat{err}(\underline{\vect{\gamma}}), & \text{when using the estimator in \cref{ch1:def:ooserror}}, \\
			\vect{\tilde{\gamma}}^{BJK}(c, \mathscr{H}^{RS}) \defeq \argmin_{\underline{\vect{\gamma}} \in \mathscr{H}^{RS}} \widetilde{err}^{BJK}(c, \underline{\vect{\gamma}}), & \text{when using the estimator in \cref{ch1:def:bjkerror}}, \\
			\vect{\tilde{\gamma}}^{AJK}(d, \mathscr{H}^{RS}) \defeq \argmin_{\underline{\vect{\gamma}} \in \mathscr{H}^{RS}} \widetilde{err}^{AJK}(d, \underline{\vect{\gamma}}), & \text{when using the estimator in \cref{ch1:def:ajkerror}}.
		\end{cases}
	\end{align*}
	
	\noindent via naive brute-force optimisation.
\end{definition}

\noindent This operation allows to keep the computational advantages of grid search, while exploring a more heterogeneous section of $\mathscr{H}$. This is especially relevant if $\vect{\gamma}$ is large, since it is difficult to generate a proper set of candidates on the basis of some deterministic or subjective rule. This formulation for the random search is rather naive. Nonetheless, \cite{bergstra2012random} showed that it is (at least) as good as more advanced versions of random search.\footnote{Further details on these algorithms can be found in \cite{solis1981minimization} and \cite{andradottir2015review}. Random search tends to be less effective for cases where the number of hyperparameters to tune is very large. For these cases, alternative techniques (e.g., simulated annealing, particle swarm optimization) surveyed in \cite{weise2009global} could help. However, since they would inevitably increase the computational burden and the complexity of the hyperparameter optimisation, they are left for future research.} It is important to stress that, when the expected error surface is flat, it is hard to pick one candidate in particular. In these cases, it is often more sensible to evaluate the whole grid of interest and use the threshold where the surface starts flattening as optimal hyperparameters.

\section{Monte Carlo study}

In this section, I report the output of a simulation study to assess the performance of the artificial jackknife on a classical time-series problem: selecting the appropriate number of lags for an autoregressive model. Namely, I focus on the bivariate causal VAR(1) process
\begin{align*}
	\vect{Y}_{t+1} = \vect{\Pi}_1 \vect{Y}_{t} + \vect{V}_{t+1},
\end{align*}

\noindent where $\vect{V}_{t+1} \widesim{w.n.} N \left(\vect{0}_{n \times 1}, \vect{\Sigma} \right)$ with $\vect{\Sigma}$ being positive definite, $t \in \mathbb{Z}$. The model is linear, easily interpretable and controlled by few parameters. As such, it serves as the most immediate use-case to benchmarks the estimators discussed above. 

I generate the model for different sizes of the sample ($T=100, 200$). The autoregressive coefficient is set to be
\begin{align*}
	\vect{\Pi}_1 &= 
	\left( 
	\begin{array}{cc}
		0.85 & -0.10 \\
		-0.10 & 0.85
	\end{array} 
	\right).
\end{align*}

\noindent These coefficients give relatively high persistent making it relatively harder to select the approriate number of lags. For simplicity, $\vect{\Sigma}$ is set to be an identity matrix and $t_0=T/2$. The missing values introduced during the subsampling process in the estimation sample are skipped. The missing values introduced following $t=t_0$ are kept to be as such. Finally, the grid employed for the selection process goes from 1 to 6 lags and, in the case of the artificial jackknife, the combinations are truncated at a maximum of 1,000 to make computations feasible.

\Cref{ch1:tab:monte_carlo_lags} reports the accuracy of the selection process associated to all error estimators, sample size and varying parametrisations. The accuracy is computed as the mean squared error between the true and estimated number of lags across 500 simulated models.\footnote{The simulations were performed on a laptop with all computations completed within a few hours.} Three key stylised facts emerge from this table. First, the jackknife-based estimators outperform the standard pseudo out-of-sample approach. Second, even removing the same number of observations, the artificial jackknife dominates the block version due to the higher number of combinations the data can be removed with this subsampling scheme. Third, the artificial jackknife dominates particularly with $T=100$ in line with \cref{ch1:proposition:finite_sample}.

	\begin{table}[!t]
	\caption{Accuracy of lag selection across error estimators, sample sizes and parameterizations.}
	\label{ch1:tab:monte_carlo_lags}
	\small
	\begin{tabularx}{\textwidth}{@{}lZZ@{}}
		\toprule 
		Error estimator & \multicolumn{2}{c}{Selection RMSE} \\
		\cmidrule(l){2-3}
		& $T=100$ & $T=200$ \\
		\midrule
		Pseudo out-of-sample error & 0.982 & 1.544 \\
		$\text{Block jackknife, }c/T= 0.1$ & 0.542 & 1.006 \\
		$\text{Artificial delete-$d$ jackknife, }d/(nT)= 0.1$ & 0.196 & 0.436 \\
		\midrule
		\bottomrule
	\end{tabularx}
	\floatfoot{\justifying This table presents the mean squared error (MSE) between the true and estimated number of lags for various error estimators under differing sample sizes and parametrizations.}
\end{table}

\section{Empirical application}\label{ch1:sec:empirics}

This section benchmarks the artificial jackknife with a a complex empirical problem: forecasting weekly exchange rates.\footnote{It is important to remark that this manuscript does not intend to find the best model (among a class of techniques) for predicting exchange rates, but rather it aims to show that the artificial jackknife is a valid approach for tuning the models in this example.}

\subsection{Penalised VARMA}\label{ch1:sec:empirics:varma}

This subsection describes the case in which forecasters form their predictions using an elastic-net VARMA($q$, $r$) models. Clearly, the following assumptions and definitions affect only the empirical example in \cref{ch1:sec:empirics}.

\begin{assumption}[VARMA model] \label{ch1:assumption:varma}
	Within \cref{ch1:sec:empirics}, forecasters form their expectations assuming that
	\begin{alignat}{2}
		&\vect{Y}_{t+1} &&= \vect{\Pi}_1 \vect{Y}_{t} + \ldots + \vect{\Pi}_q \vect{Y}_{t-q+1} + \vect{\Xi}_1 \vect{V}_{t} + \ldots + \vect{\Xi}_r \vect{V}_{t-r+1} + \vect{V}_{t+1}, \label{ch1:eq:varma}
	\end{alignat}
	
	\noindent where $\vect{V}_{t+1} \widesim{w.n.} N \left(\vect{0}_{n \times 1}, \vect{\Sigma} \right)$ with $\vect{\Sigma}$ being positive definite, $t \in \mathbb{Z}$.\footnote{The data is assumed to have zero mean and unit standard deviation for simplicity of notation.} The autoregressive and moving average coefficients are $n \times n$ matrices for which the VARMA is causal and invertible \citep[pp. 418-420]{brockwell1991time}.
\end{assumption}

\noindent For simplicity of notation, let
\begin{alignat*}{2}
	&\vect{\Pi} &&\defeq \begin{pmatrix} \vect{\Pi}_{1} & \ldots & \vect{\Pi}_{q} \end{pmatrix}, \\
	&\vect{\Xi} &&\defeq \begin{pmatrix}\vect{\Xi}_{1} & \ldots & \vect{\Xi}_{r} \end{pmatrix}.
\end{alignat*}

\noindent Moreover, I consider only parametrisations where $\min(q,r)=0$. Under \cref{ch1:assumption:lags}, $\max(q,r) > 0$. Therefore, letting $\min(q,r)=0$ does not exclude the white noise case, since the model could be parametrised to have autoregressive and moving average coefficients equal to zero. This point is purely to simplify the notation in the Online Appendix.

	\begin{definition}[Penalised maximum likelihood estimation] \label{ch1:def:varma_estimation}
		Forecasters use penalised maximum likelihood estimation to estimate the estimated VARMA coefficients. With complete data, this implies
		\begin{align*}
			\vect{\hat{\theta}}_{s}(\vect{\gamma}) \defeq \argmax_{\underline{\vect{\theta}} \, \in \, \mathscr{R}} \, \mathcal{L}(\underline{\vect{\theta}} \,|\, \vect{Y}_{1:s}) - \mathcal{P}(\underline{\vect{\theta}}, \vect{\gamma}),
		\end{align*}
		
		\noindent where $\mathscr{R}$ is the region of interest for the parameters implicitly defined in \cref{ch1:assumption:varma},
		\begin{align*}
			\mathcal{L}(\underline{\vect{\theta}} \,|\, \vect{Y}_{1:s}) \simeq - \frac{s}{2} \ln | \underline{\vect{\Sigma}} | - \frac{1}{2} \Tr \left[\sum_{t=1}^{s} \underline{\vect{\Sigma}}^{-1} \vect{V}_{t}(\underline{\vect{\theta}}) \vect{V}_{t}(\underline{\vect{\theta}})'\right]
		\end{align*}
		
		\noindent denotes the log-likelihood of the VARMA model \citep[][ch. 11]{lutkepohl2005new} and $\mathcal{P}(\underline{\vect{\theta}}, \vect{\gamma})$ is a penalty function, for $\max(q,r) \leq s \leq T$.\footnote{The innovations $\vect{V}_{t}(\underline{\vect{\theta}}) \equiv \vect{V}_{t}$ in \cref{ch1:eq:varma}. This notation is used for stressing its dependence from the coefficients in $\underline{\vect{\theta}}$ and obtain a compact formula.} By extension, $\underline{\vect{\Pi}}$, $\underline{\vect{\Xi}}$ and $\underline{\vect{\Sigma}}$ are the VARMA coefficients built from $\underline{\vect{\theta}}$. 
	\end{definition}

	\noindent Performing penalised maximum likelihood with incomplete data is non-trivial. In order to overcome the related complexities, this article uses an Expectation-Conditional Maximisation (ECM) algorithm \citep{meng1993maximum}. The details of this iterative estimation procedure are described in the Online Appendix.
	
	The penalty function of interest for this empirical application builds on the elastic-net literature \citep{zou2005regularization, zou2009adaptive}.
	
	\begin{definition}[Generalised elastic-net penalty] \label{ch1:def:elasticnet_pure}
		For any $p \in \mathbb{N}$, let
		\begin{align*}
			&\vect{\Gamma}(\vect{\gamma}, p) \defeq \lambda \begin{pmatrix}
				\vect{I}_{n} &\vect{0}_{n \times n} &\ldots &\vect{0}_{n \times n} \\
				\vect{0}_{n \times n} &\beta \cdot \vect{I}_{n} &\ldots &\vect{0}_{n \times n} \\
				\vdots &\ddots &\ddots &\vdots \\
				\vect{0}_{n \times n} &\ldots &\ldots & \beta^{p-1} \cdot \vect{I}_{n} \end{pmatrix}
		\end{align*}
		
		\noindent where $\vect{\gamma}\defeq(q\;\,r\;\, \lambda\;\, \alpha\;\, \beta)'$ is a given vector of  hyperparameters with $\lambda \geq 0$, $0 \leq \alpha \leq 1$ and $\beta \geq 1$. Building on that, this manuscript uses the penalty
		\begin{align} \label{ch1:eq:elasticnet_pure}
			\mathcal{P} (\underline{\vect{\theta}}, \vect{\gamma}) \defeq \begin{cases}
				\frac{1-\alpha}{2} \big\Vert \underline{\vect{\Pi}} \, \vect{\Gamma}(\vect{\gamma}, q)^{\frac{1}{2}} \big\Vert_{\text{F}}^2 +\frac{\alpha}{2} \big\Vert \underline{\vect{\Pi}} \, \vect{\Gamma}(\vect{\gamma}, q) \big\Vert_{1,1} & \text{if } q > 0 \text{ and } r=0, \\
				\frac{1-\alpha}{2} \big\Vert \underline{\vect{\Xi}} \, \vect{\Gamma}(\vect{\gamma}, r)^{\frac{1}{2}} \big\Vert_{\text{F}}^2 +\frac{\alpha}{2} \big\Vert \underline{\vect{\Xi}} \, \vect{\Gamma}(\vect{\gamma}, r) \big\Vert_{1,1} & \text{if } q=0 \text{ and } r > 0.
			\end{cases}
		\end{align}
	\end{definition}

	\begin{remark}
		Note that when the penalty is active (i.e., $\lambda > 0$), $\vect{\Gamma}(\vect{\gamma}, q)$ and $\vect{\Gamma}(\vect{\gamma}, r)$ are diagonal and positive definite matrices, and thus
		\begin{align*}
			\mathcal{P} (\underline{\vect{\theta}}, \vect{\gamma}) = \begin{cases}
				\sum_{i=1}^n \sum_{j=1}^{nq} \frac{1-\alpha}{2} \,\underline{\Pi}_{\,i,j}^2 \, \left[\vect{\Gamma}(\vect{\gamma}, q) \right]_{j,j} + \frac{\alpha}{2} \, |\underline{\Pi}_{\,i,j}| \, \left[\vect{\Gamma}(\vect{\gamma}, q) \right]_{j,j} & \text{if } q > 0 \text{ and } r=0, \\
				\sum_{i=1}^n \sum_{j=1}^{nr} \frac{1-\alpha}{2} \,\underline{\Xi}_{\,i,j}^2 \, \left[\vect{\Gamma}(\vect{\gamma}, r) \right]_{j,j} + \frac{\alpha}{2} \, |\underline{\Xi}_{\,i,j}| \, \left[\vect{\Gamma}(\vect{\gamma}, r) \right]_{j,j} & \text{if } q=0 \text{ and } r > 0.
			\end{cases}
		\end{align*}
	\end{remark}
	
	\noindent The penalty $\mathcal{P} (\underline{\vect{\theta}}, \vect{\gamma})$ is a generalisation of the elastic-net that allows to penalise more autoregressive and moving average coefficients  referring to distant points in time. 
	
	As for its standard implementation, when $\alpha=1$ and $\alpha=0$ the function is equivalent to the LASSO \citep{tibshirani1996regression} and ridge \citep{hoerl1970ridge} penalties. These penalties perform differently depending on the empirical setting in which they are employed. LASSO gives a sparse representation of the model and thus a simple regression in few predictors. Ridge does not select subsets of regressors, but it shrinks all of them jointly. The ability of LASSO in selecting the same covariates over time is rather poor when some of them are highly correlated. This is extensively described in \cite{zou2005regularization}.
	 
	 For $0 < \alpha < 1$ the model allows for a sparse model and benefits from the co-movement of correlated predictors. With respect to the standard elastic-net, the penalty function in \cref{ch1:eq:elasticnet_pure} includes $\beta$, an additional hyperparameter. If $\beta > 1$, then $\mathcal{P}(\underline{\vect{\theta}}, \vect{\gamma})$ penalises more coefficients referring to distant points in time. This idea is commonly used in time series and a simple parallel can be made by looking at Bayesian VARs with Minnesota priors \citep{doan1984forecasting, litterman1986forecasting}. Indeed, in stationary settings, this set of priors shrinks the vector autoregression toward a white noise (i.e., it shrinks the coefficients to zero) and penalises more distant lags. The penalty in \cref{ch1:eq:elasticnet_pure} is similar in spirit, but it allows for a sparse representation of the model and for the use of moving average coefficients.

	\begin{figure}[!t]
		\begin{subfigure}[t]{\textwidth}
			\centering
			\includegraphics[width=\textwidth]{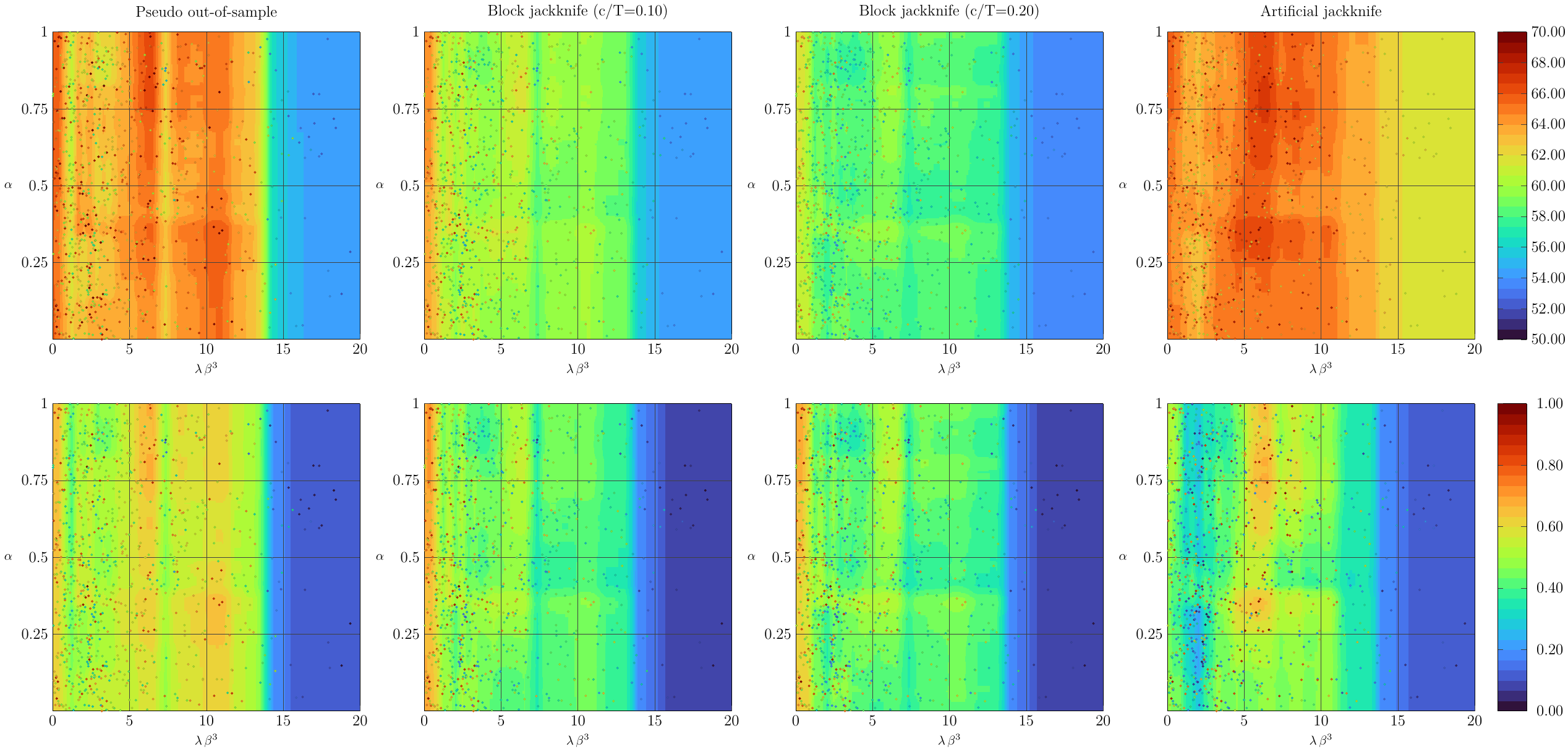}
			\caption{Vector autoregression.}
		\end{subfigure}
		
		\begin{subfigure}[t]{\textwidth}
			\centering
			\vspace{8pt}
			\includegraphics[width=\textwidth]{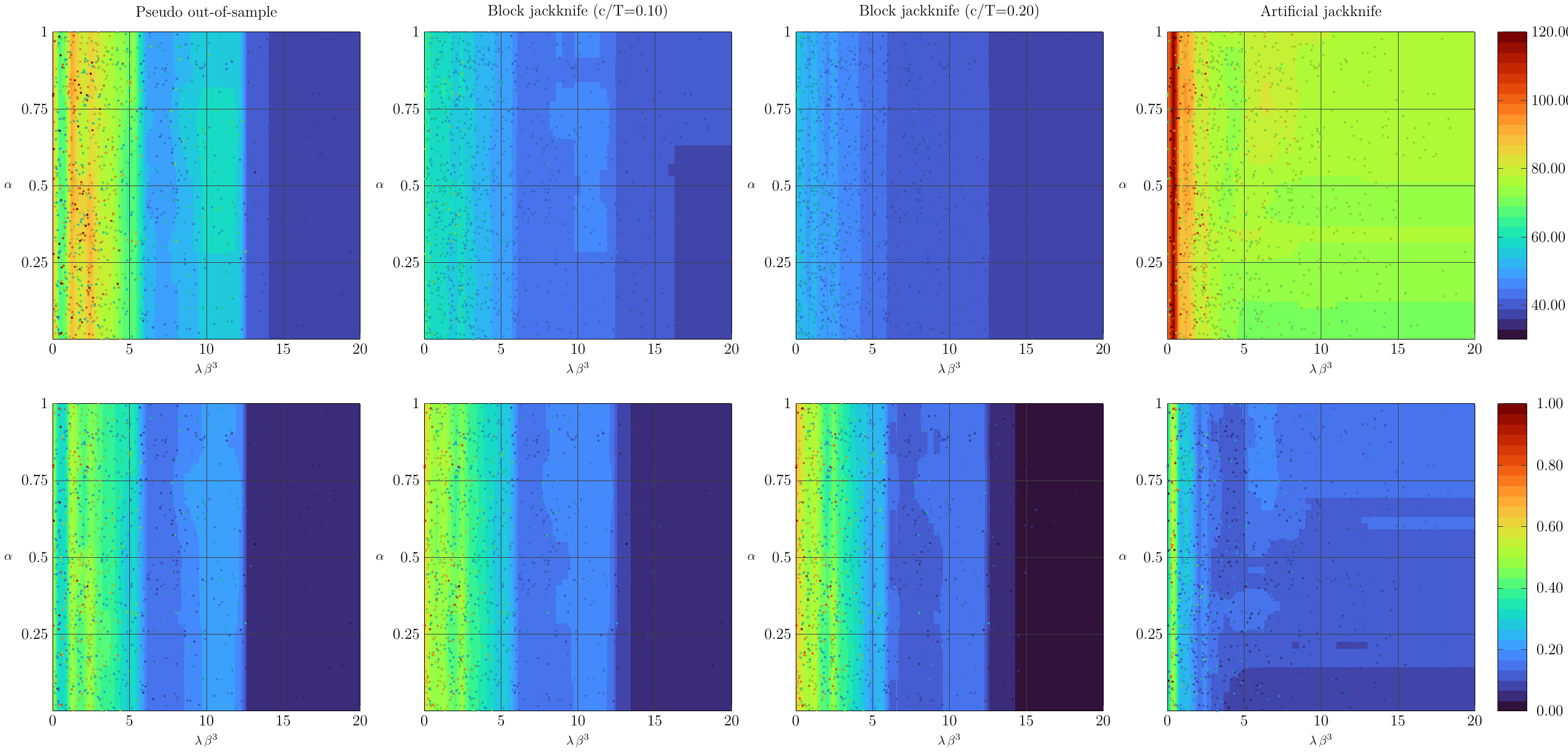}
			\caption{Vector moving average.}
		\end{subfigure}
		
		\caption{Estimated errors for the candidate hyperparameters in $\mathscr{H}$.}
		\label{ch1:fig:contours}
		\floatfoot{\justifying For each model and subsampling method, the first row describes the estimated error in absolute terms, while the second one shows it in relative terms. $\lambda\,\beta^{3}$ denotes the shrinkage associated to the farthest lag. The block jackknife output is adjusted to reduce the finite-sample methodological defects as described at the start of \cref{ch1:sec:methodology:novel_estimators}.}
	\end{figure}
	
	\subsection{Results} \label{ch1:sec:empirics:results}
	
	The time series for the exchange rates are collected from the Federal Reserve Board H.10 and include regular weekly (Friday, EOP) observations from the first week of January 1999 to the last week of December 2020, for a set of major economies reported in the Online Appendix. This dataset contains a total of 1,148 weeks and 21,812 observations. These are all the exchange rates in the Federal Reserve Board H.10 that did not have a fixed or pegged rate with the dollar in the sample. Moreover, these exchange rates are not taken in levels, but they are transformed in weekly log-returns instead.
	
	The sample is divided into three blocks: a presample (January 1999 to December 1999), a selection sample (January 2000 to December 2001) and a test sample (January 2002 to December 2020). The presample is used for computing $\vect{w}$ and then discarded. Each entry in this vector is equal to the reciprocal of the variance of the corresponding series in the presample. This ensures that each series is weighted equally, regardless of its volatility. The grid of candidate hyperparameters $\mathscr{H} = \mathscr{H}_p \times \mathscr{H}_\lambda \times \mathscr{H}_\alpha \times \mathscr{H}_\beta$ is explored via random search over the selection sample using, in turn, each of the error estimators described above. This is done settings $\mathscr{H}_p \defeq \{4\}$, $\mathscr{H}_\lambda \defeq [10^{-2}, 2.5]$, $\mathscr{H}_\alpha \defeq [0, 1]$ and $\mathscr{H}_\beta \defeq [1, 2]$ and letting the methods based on pseudo out-of-sample criteria defining $t_0$ to be such that the part of the selection sample used for estimation purposes ends in December 2000. The set $\mathscr{H}_p$ fixes the number of lags to 4: a value considered large enough to forecast the weekly financial returns.\footnote{Since the vector moving averages in this manuscript are constrained to be invertible, they account for a higher persistence than the vector autoregressions of the same order. Indeed, any invertible VMA can be equivalently thought as a VAR($\infty$).} The sets referring to the remaining hyperparameters allow to control the overall shrinkage level and kill superfluous lags, if needed. To assess the accuracy of each estimator, I compute the mean squared difference between these estimates and the realised pseudo out-of-sample error over the test sample.\footnote{The realised pseudo out-of-sample error are based on a model estimated over the full selection sample.} The smaller the difference, the higher the precision of the estimator.
	
	\Cref{ch1:fig:contours} describes the random search output. The vector autoregression results show a series of important features. First, it is evident that the estimated error decreases as the shrinkage level increases, regardless of the estimator. Second, the area with the lowest estimated error is where $\lambda \beta^3 \geq 15$. This location seems independent from $\alpha$ and estimator. Third, there are strong differences in the scale of the estimates obtained via different estimators. Indeed, the artificial jackknife estimates are the most conservative, since the figure is higher in scale across all candidate hyperparameters. The pseudo out-of-sample gives similar estimates for any configuration with $\lambda \beta^3 < 15$, but a more pronounced fall before the common flattening. The block jackknife measures are the least conservative. The picture observed through the lenses of the vector moving average results is quite different. Indeed, while the estimated error still decreases when the shrinkage level increases. The artificial jackknife estimates decrease sharply for much lower $\lambda \beta^3$ than the other estimators. Furthermore, they start flattening at a much smaller shrinkage level compared to the benchmarks.
	
	\begin{table}[!t]
		\caption{Estimators evaluation.}
		\label{ch1:tab:selection_rmse}
		\small
		\begin{tabularx}{\textwidth}{@{}lZZ@{}}
			\toprule 
			Error estimator & \multicolumn{2}{c}{Selection RMSE} \\
			\cmidrule(l){2-3}
			& Vector autoregression & Vector moving average \\
			\midrule
			In-sample error &  7.28 & 1.50 \\
			Pseudo out-of-sample error & 1.00 & 1.00 \\
			$\text{Block jackknife, }c/T= 0.1$ & 1.68 & 1.20 \\
			$\text{Block jackknife, }c/T= 0.2$ & 2.47 & 1.30 \\
			$\text{Block jackknife, }c/T= 0.1$ (adjusted) \hspace{1em} & 1.17 & 1.14 \\
			$\text{Block jackknife, }c/T= 0.2$ (adjusted) & 1.28 & 1.20 \\
			Artificial delete-$\hat{d}$ jackknife & 0.93 & 0.89 \\
			\bottomrule
		\end{tabularx}
		\floatfoot{\justifying The performance of the estimators is measured using the average squared differences between the estimated errors on the selection sample (January 2000 to December 2001) and the realised pseudo out-of-sample errors on the test sample (January 2002 to December 2020). The average is computed across all candidate hyperparameters. Note that the values are scaled, with numbers below one indicating better performance compared to the pseudo out-of-sample error estimator.}
	\end{table}
	
	As anticipated at the start of this subsection, the estimates are evaluated against the realised pseudo out-of-sample error over the test sample. \Cref{ch1:tab:selection_rmse} summarizes this comparison for both forecasting models. Instead of reporting the raw squared differences between the estimated and realised errors, the results are scaled such that values below one indicate superior performance relative to the pseudo out-of-sample error estimator. These results show that the artificial jackknife gives the best results. This is evident both for vector autoregressions and vector moving averages. A further interesting result is that the pseudo out-of-sample is better than the block jackknife (both raw and adjusted). This is likely due to the small number of partitions that the block jackknife is able to generate in this empirical application. Finally, it follows from \cref{ch1:tab:selection_rmse} and \cref{ch1:fig:contours} that the best configuration for vector moving averages requires a much smaller shrinkage level compared to vector autoregressions.

\section{Concluding comments}\label{ch1:sec:conclusions}

This article proposes a new approach for selecting hyperparameters in time series denoted as artificial delete-$d$ jackknife: a generalisation of the delete-$d$ jackknife.

By contrast with existing approaches, the artificial delete-$d$ jackknife can partition dependent data into a large set of unique partitions, even when $T$ is relatively small. These partitions are used for constructing a robust forecast error estimator, based on pseudo out-of-sample evaluations. The artificial delete-$d$ jackknife has strong finite-sample advantages and converges in probability to the true error. Empirical results on weekly exchange rate returns are also promising. It also outperforms in-sample estimators and, consequently, information criteria, such as the AIC, that rely on them.

While the theory developed in this paper is based on a weighted mean square loss, the artificial jackknife error could be extended to other loss functions for prediction and classification problems. Also, it could be expanded to compute the uncertainty around sample statistics in time series. These and a few other points are not fully developed in this article and they are left for future research.

% Notes at the end
%{
%	\titlespacing\section{0pt}{12pt plus 4pt minus 2pt}{0pt plus 2pt minus 2pt}
%	\setlength{\parskip}{1.5ex}
%	\theendnotes
%}

% References
\bibliographystyle{abbrvnat}
\bibliography{./biblio}

% Save counters for assumptions, corollaries, equations, figures, propositions and tables to a file called "counters.tex"
% This is used to cross-reference paper.texs

% Generate file
\newoutputstream{stream}
\openoutputfile{counters}{stream}

% Add to "counters.tex"
\addtostream{stream}{
	\protect\setcounter{proposition}{\arabic{proposition}}
	\protect\setcounter{assumption}{\arabic{assumption}}
	\protect\setcounter{definition}{\arabic{definition}}
	\protect\setcounter{remark}{\arabic{remark}}
	\protect\setcounter{equation}{\arabic{equation}}
}

% Close file
\closeoutputstream{stream}

\end{document}